\documentclass[%
 reprint,
 amsmath,amssymb,
 aps,
]{revtex4-2}
\usepackage{chemformula}
\usepackage{graphicx}
\usepackage{dcolumn}
\usepackage{bm}
\usepackage{amsmath}
\usepackage[version=4]{mhchem}

\begin{document}

\preprint{APS/123-QED}

\title{Rotationally Resolved Spectroscopy of a Single Polyatomic Molecule}

\author{Aaron Calvin}
\affiliation{Department of Physics, University of California, Santa Barbara}%

\author{Merrell Brzeczek}
\affiliation{Department of Physics, University of California, Santa Barbara}

\author{Samuel Kresch}
\affiliation{Department of Physics, University of California, Santa Barbara}

\author{Elijah Lane}
\affiliation{Department of Physics, University of California, Santa Barbara}

\author{Lincoln Satterthwaite}
\affiliation{Department of Physics, University of California, Santa Barbara}

\author{Desi Hawkins}
\affiliation{Department of Physics, University of California, Santa Barbara}

\author{David Patterson}
\affiliation{Department of Physics, University of California, Santa Barbara}
\email{davepatterson@ucsb.edu}

\date{\today}

\begin{abstract}

We report the rotationally resolved spectrum of a single polyatomic molecular ion in the gas phase. Building upon the recently developed inelastic recoil spectroscopy (IRS) technique, we have achieved a spectral resolution sufficient to observe resolved rotational-vibrational transitions of a trapped cyclopropenyl cation, c-C$_{3}$H$_{3}^+$. The high precision of IRS shown in this work presents an attractive platform for astrochemistry, single molecule chiral detection, and tests of fundamental physics. 

\end{abstract}

\maketitle

Infrared spectroscopy is a vital tool for both chemical analytics and fundamental science. Molecules provide enhanced sensitivity to certain exotic interactions compared to atoms and present new opportunities in testing fundamental physics \cite{Hutzler2020, DeMille2024}. The complexity of polyatomic molecules poses a significant challenge in precision measurement. The ideal spectroscopic technique would combine high resolution, single molecule sensitivity, and general applicability. To date no single method has simultaneously satisfied these requirements.

Quantum logic spectroscopy methods demonstrate single molecule sensitivity and superlative resolution\cite{Sinhal2020, chou2020frequency, Collopy2023, Markus2024, Wolf2024} but to date are limited to diatomic molecular ions. In contrast, action spectroscopy techniques such as the newly developed Leak Out Spectroscopy (LOS) yield excellent resolution and generality but lose some resolution as compared to QLS\cite{Schmid2022, Asvany2023, Gupta2023, Bast2023, Silva2023b, Steenbakkers2024, Schlemmer2023, Silva2024}. Briefly, Leak Out Spectroscopy relies on vibrational to translational (V-T) energy transfer during inelastic collisions with a buffer gas particle (typically He, Ne, or N$_2$) to selectively eject vibrationally excited molecular ions from an ion trap into a detector, where they are destroyed. The loss rate as a function of infrared frequency yields a high resolution spectrum with few limitations on the molecular ion species, but the destructive detection of ejected ions precludes the use of this technique as a single molecule tool. 

A straightforward modification of LOS using molecular ions co-trapped with laser-cooled atomic ions allows non-destructive detection of inelastic recoil events to obtain a spectrum of a single molecule. Termed Inelastic Recoil Spectroscopy (IRS), the spectra of several molecular ion species was obtained at the single molecule level \cite{CalvinPRA2023}. IRS relies on the same inelastic recoil events as LOS, but in the case of IRS the events do not eject the ion from the trap, but instead reconfigure a Coulomb crystal containing the molecular ion of interest and one or more laser-cooled atomic ions. The use of a Coulomb crystal has benefits for precision spectroscopy, as the molecule is significantly motionally cooled.  This reduces Doppler broadening, and cooling to the Lamb-Dicke regime where the first-order Doppler effect is suppressed is expected to be straightforward. Detection of Coulomb crystal reconfiguration due to vibrational-to-translational energy transfer can be achieved by observing the molecular ion hopping between separate trapped ensembles separated by a potential barrier, or, by observing the Coulomb crystal briefly melting after an inelastic collision event. Both implementations previously published have low resolution spectra due to a wide-bandwidth mid-IR source\cite{CalvinPRA2023}. 

Here, we extend IRS to measure a high precision spectrum of a single polyatomic molecule (c-C$_3$H$_3^+$). Non-destructive high resolution infrared spectroscopy of a single molecule is a critical prerequisite for single molecule chiral readout, and an important stepping stone towards ultra-precise spectroscopy and the search for parity violation in chiral molecules\cite{Eduardus2023}.

As described in figure 1, we detect inelastic recoil events via Coulomb crystal reconfiguration. A c-C$_3$H$_3^+$ molecule which is vibrationally excited in the 3130 cm$^{-1}$ C-H stretch band has 4513 K (389 meV) internal energy. The molecular ion has a maximum of 419.8 K (36 meV)  or 1530 K (132 meV) kinetic energy upon a quenching collision with He or Ne respectively, although redistribution of energy among vibrational modes likely results in a significantly lower kinetic energy. This increase in kinetic energy from a quenching collision is sufficient to briefly melt the Coulomb crystal, which has a temperature in the mK range, but insufficient to fully eject the molecular ion from the trap. The laser cooled atomic ion quickly re-cools ($\lesssim$ 100 ms) forming a crystal of pure Sr$^+$ while the molecular ion cools via buffer gas collisions and ion-ion Coulomb interactions on a timescale of about 200 ms. We detect these inelastic recoil events by monitoring changes in the Coulomb crystal's combined mass through non-destructive mass detection. 

A single analyte molecular ion and a small number (typically one-three) Sr$^+$ ions are co-loaded into an ion trap, as described in Ref. \cite{CalvinPRA2023} and shown in Fig. 1a.  The ensemble is motionally cooled via laser cooling the Sr$^+$ ions on the $5s ^2S_{1/2}  \rightarrow 5p ^2P_{1/2} $  transition at 422 nm (Doppler temperature $\sim$300 $\mu$K).  Cold (10 K) He is added at a pressure of 5$\times 10^{-7}$ - 1$\times 10^{-6}$ torr above the estimated background He pressure of $\sim$10$^{-8}$ torr.  The incoming He is cooled to 10 K via thermalization with a copper tube thermally anchored to the cryogenic housing of the trap. The He serves as both an inelastic collision partner and thermalizes the rotational state of the molecular ion to $\sim$10 K. At this temperature, the molecule occupies $\sim$50 rotational states and is randomly redistributed among these states at the collision frequency of $\sim$400 Hz. 

In the optically saturated regime in which we operate, we expect the molecule to be excited whenever it is randomly scattered into relevant initial state, yielding an excitation rate of 8 s$^{-1}$. As shown in Fig. 1(d), we detect inelastic collision events at a  slower rate of $\sim$0.4 s$^{-1}$. This is in part because not every vibrationally excited molecule undergoes an inelastic collision before it radiatively decays back to the ground state, and in part because not every inelastic collision event is detected. Details depend on the molecule and density of states, but the inelastic collision rate is typically some fraction of the elastic collision rate. The similar sized molecular ion \ch{C2H3+} requires $\sim$100 collisions with \ch{H2} to undergo an inelastic collision\cite{Markus2020}. The higher density of states in larger molecules is expected to result in a larger inelastic collision rate, at the cost of lower kinetic energy imparted to the heavier molecule by a collision with He.   Although we observe inelastic recoil events with He for this relatively light molecular ion, we find a small amount of Ne improves the number of inelastic events recorded. 

Inelastic collisions compete with elastic collisions with He, as only a few elastic collisions with background He may cool a kinetically excited molecule back to the gas temperature of $\sim$10 Kelvin, from which it can rapidly cool and re-join the Coulomb crystal via ion-ion interactions\cite{Wester2009, Bast2023}. The He buffer gas pressure of $\sim$1$\times 10^{-6}$ torr is chosen to be sufficient to produce recoil events, but not so high as to recool the crystal before it can be detected, or to melt the crystal entirely due to elastic collisions between trapped ions and He gas atoms. Laser-cooling parameters, trap parameters, and the amount of buffer gas tune the sympathetic re-cooling time of the molecular ion and provide a rich parameter space to achieve balance between efficiently inducing inelastic recoils and the deleterious quenching of the recoil signal caused by a collision of a kinetically excited molecular ion with a He atom.

The mid-IR source (DLC TOPO - Toptica Photonics) is co-aligned with the cooling lasers and sent axially through the trap with a beam waist diameter of 0.5 mm. The infrared frequency is measured with a commercial wavemeter (Bristol 671A-MIR) with an accuracy of 30 MHz (0.001 cm$^{-1}$).  The trapped ensemble is illuminated with tunable infrared light in the 3130 cm$^{-1}$ (CH stretch) band. When this light is tuned to be resonant with a ro-vibrational transition associated with a single initial rotational state, the molecule will be vibrationally excited, and then momentarily ejected from the Coulomb crystal, but not the trap, as described above.  These ejection events are detected via the fluorescence of the co-trapped atomic ions, as shown in Fig. 1(a)-(c).  Scanning the infrared laser while monitoring these ejections events yields a spectrum (Fig. 2 and Fig. 3). 

\begin{figure}
    \centering
    \includegraphics[width = 0.5\textwidth]{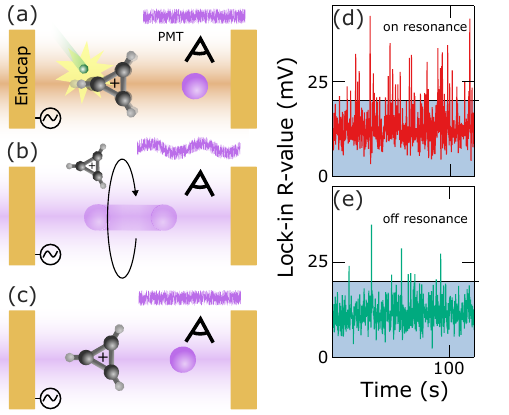}
    \caption{
\textbf{Non-destructive inelastic recoil detection. }(a) a molecular ion is co-trapped and sympathetically cooled with a laser-cooled Sr$^+$. The ions are confined radially in a linear quadrupole (not shown) and axially by two endcaps. A constant oscillating voltage (tickle) is applied to an endcap at the Sr$^+$ secular frequency. No excitation of the mixed-species crystal occurs due to the mismatch of the mass of the crystal to the applied tickle frequency. However, upon ro-vibrational excitation by a mid-IR source and subsequent inelastic collision, the molecular ion is ejected from the Coulomb crystal but still remains in the trap (b). The Sr$^+$ motion is amplified by the now-on-resonance tickle and the fluorescence is modulated at the secular frequency. (c) The molecular ion eventually re-cools into the crystallized configuration. The fluorescence modulation drops and the measurement is repeated.  A lock-in amplifier detects the Sr$^+$ fluorescence modulation. (d) With the mid IR source on resonance (3134.7278 cm$^{-1}$), multiple inelastic collision events are detected compared to off resonance (3134.7255 cm$^{-1}$) (e). The time the lock in R-value spends above a pre-determined threshold (20 mV) is used as the signal for the spectra reported here. 
}
    \label{fig:flippytoon}
\end{figure}

\begin{figure}
    \centering
    \includegraphics[width = 0.5\textwidth]{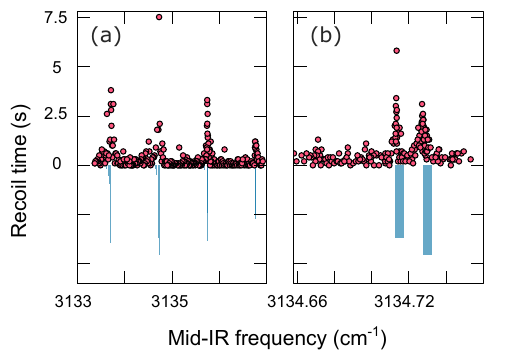}
    \caption{\textbf{Power-broadened spectra.} (a) A broadened, partially rotationally resolved spectrum shows higher probability of recoil event occurrences when the mid-IR source is tuned near resonance to a sub-manifold. Data points from this measurement are displayed as a scatter plot. Peak positions from previous measurements\cite{Marimuthu2020, Schlemmer2023PrivateComm} are displayed as solid lines with relative heights scaled to match the expected population of the initial rotational state at 10K. The four features are enough to distinguish this m/z = 39 molecule as c-C$_{3}$H$_{3}^+$. (b) Each band from (a) has a sub-manifold of rotational levels that is fully resolved with lower mid-IR intensity. }
    \label{fig:spec1}
\end{figure}

A fast but modest resolution spectrum (Fig. 2 (a)) can be obtained by increasing the infrared power to a relatively high intensity of 3 $\times 10^{6}$ W/m$^2$ and scanning the infrared frequency in fairly coarse steps (350 MHz, 0.012 cm$^{-1}$) between data points.  Under these conditions, the width of the spectral feature is verified to be more than the frequency step to avoid missing spectral features. An estimated\cite{Levine2012} linewidth of $\sim$1000 MHz (0.033 cm$^{-1}$)  for this high intensity scan qualitatively agrees with the observed linewidths given uncertainties in the intensity due to beam alignment and OPO power drifts over the course of the scan. The resulting spectrum has a structure that is consistent with previous measurements of the c-C$_{3}$H$_{3}^+$ isomer\cite{Marimuthu2020, Schlemmer2023PrivateComm}.

Each of the four bands in Fig. 2(b) has further structure, which we resolve with modest mid-IR intensity ($\sim$3 $\times 10^{4}$ W/m$^2$) as shown in Fig. 3(b) . The fully rotationally resolved $\nu_{0}, J'' = 2, K'' = 1 \rightarrow  \nu_{1}, J' = 1, K' = 0   $ (left peak) and $\nu_{0}, J'' = 2, K'' = 2 \rightarrow  \nu_{1}, J' = 3, K' = 3   $  (right peak) transitions are clearly observed. For the beam intensity, we estimate a 100 MHz (0.0035 cm$^{-1}$) linewidth which qualitatively agrees with the observed linewidth.

In order to show the potential for high precision measurement, we further reduce the intensity of the mid-IR OPO beam to about 1.5 $\times 10^{3}$ W/m$^2$ using neutral density filters. As shown in Fig. 3, we are able to measure the rightmost transition in Fig. 2(b) with  linewidth of $\sim$ 23 MHz which is limited by the resolution of the wavemeter. 

\begin{figure}[ht!]
    \centering
    \includegraphics[width = 0.5\textwidth]{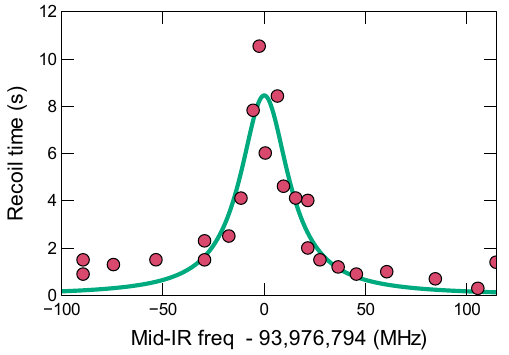}
    \caption{\textbf{Narrow linewidth spectrum}. The measurement of the $\nu_{0}, J'' = 2, K'' = 2 \rightarrow  \nu_{1}, J' = 3, K' = 3   $ at much lower mid-IR intensity yields a spectrum with a resolution limited by the resolution of the wavemeter. A linewidth of 23 MHz is estimated with a Lorentzian fit of the data.}
    \label{fig:spec1}
\end{figure}

The measurement of ro-vibrationally resolved resonances to a linewidth limited to our wavemeter represents a significant advance toward high precision measurements with a single polyatomic molecule. In particular, double resonance MW-IR spectroscopy was previously shown for an ensemble of molecular ions using the closely related LOS method\cite{Asvany2023}. This method allows for direct measurement of extremely narrow microwave-frequency rotational transitions, with no requirement of an ultra-narrow infrared light source. This method uses mm-wave radiation to drive a pure rotational transition, which de-populates the initial rotational state of a ro-vibrational line excited by the mid-IR source. Resonance of the mm-wave source is thus detected as a depletion of the inelastic recoil signal. The c-C$_{3}$H$_{3}^+$ molecule presented here has no dipole moment for pure rotational transitions, however many other candidate molecules are available. 

Our non-destructive implementation of IRS at the single molecule level would enable high precision measurement of rotational transitions for single molecular ions trapped in the Lamb-Dicke regime, where the molecule is confined to a region smaller than the wavelength of probe light. Under conditions demonstrated in this work, the molecule is in the Lamb-Dicke regime for microwave transitions but not infrared transitions. A tighter trap, with higher secular frequencies, would be used for spectroscopy of mid-IR transitions in the Doppler-free Lamb-Dicke regime. Ro-vibrational transitions in this regime would be natural-linewidth limited, with linewidths on the order of kHz. The current implementation of this technique would collisionally broaden transitions on the order of a few kHz.

We produced the molecular ions in this study from a photo-fragmentation reaction. In this case, a single C$_{5}$H$_{5}^+$ molecular ion is trapped directly from electron impact-ionized toluene using the setup described in Ref. \cite{Calvin2023, Eierman2023}. This molecular ion is co-trapped and sympathetically cooled with the desired number of laser-cooled Sr$^+$ atoms. We illuminate this molecule with a broad-linewidth OPO, as in Ref. \cite{CalvinPRA2023}.  A vibrational resonance can be seen around 3025 cm$^{-1}$. After several seconds of exposure to 3025 cm$^{-1}$ radiation, the C$_{5}$H$_{5}^+$ fragments, leaving C$_{3}$H$_{3}^+$ trapped with the Sr$^+$ ions. The mass of both the precursor and the fragment are verified via measurement of the axial secular frequency. The C$_{3}$H$_{3}^+$ ion has two isomers, linear and cyclic.  We unambiguously verify that the   c-C$_{3}$H$_{3}^+$ isomer is produced from this photofragmentation from the rotationally-resolved resonances spectra consistent with this species \cite{Schlemmer2023PrivateComm, Marimuthu2020}.

High precision measurements of chiral molecular ions is a promising avenue for detecting parity-violating effects in molecules \cite{Eduardus2023}. One challenging aspect of this search is the requirement for enantiopure samples\cite{Daussy1999, Darqui2010}. Although enantiopure samples of biogenic and small organic molecules are readily available, relatively exotic chiral molecular ions with heavy nuclei are expected to have a larger parity violating shift of vibrational resonances\cite{Quack2022, Laerdahl1999, Schwerdtfeger2003}. The enantiospecific chemistry of these compounds is lacking, complicating the search. As a single molecule is inherently enantiopure, the problem of preparing an enantiopure sample is reduced to the problem of determining which enantiomer is trapped. This problem has been solved with chiral three wave mixing \cite{Patterson2013} on microwave transitions in large ensembles of cold neutral molecules. Implementation to determine chirality of a single molecule in an ion trap is a straightforward extension of the double resonance spectroscopy described above. 

Here we have demonstrated a non-destructive spectroscopic technique which fully resolves rovibrational transitions in a single polyatomic molecule. We use this method to verify c-C$_{3}$H$_{3}^+$ as a product of the photofragmentation of C$_{5}$H$_{5}^+$. The expected generality of this method naturally lends itself as an analytical technique in situations in which the molecular ion of interest is produced in insufficient quantity to study using destructive methods. Anticipated extensions of this technique include high resolution double-resonance spectroscopy of pure rotational transitions and single molecule chirally-sensitive spectroscopy. This platform enables the high precision measurement of isolated molecules in the Lamb-Dicke regime, with the tantalizing prospect for precision measurement of parity violation in chiral molecules from a racemic mixture of a molecular candidate. 

 \vspace{5mm} 

This work has been supported by the US National Science Foundation (NSF CHE-1912105) and the Air Force Office of Scientific Research (MURI FA9550-20-1-0323).

\end{document}